\documentclass[aps,prd,preprint,superscriptaddress,amsfonts,amssymb,amsmath,nofootinbib,article]{revtex4-1}
\usepackage{graphicx,bm,color,ulem,latexsym}
\usepackage[colorlinks,linkcolor=magenta,anchorcolor=blue,citecolor=blue]{hyperref}
\usepackage{appendix}
\usepackage{makecell}
\usepackage{graphicx}
\usepackage{float}

\begin{document}

\title{Parity violation in framework of nonmetricity gravity}

\author{Zhiyuan Yu}
\email{d202280102@hust.edu.cn}
\affiliation{School of Physics, Huazhong University of Science and Technology\\
Wuhan, 430074, China}

\author{Zhengsheng Yang}
\email{zhengshengy@hust.edu.cn}
\affiliation{School of Physics, Huazhong University of Science and Technology\\
Wuhan, 430074, China}

\author{Taotao Qiu}
\email{qiutt@hust.edu.cn}
\thanks{Corresponding author.}
\affiliation{School of Physics, Huazhong University of Science and Technology\\
Wuhan, 430074, China}

\begin{abstract}
The latest observational data of Planck satellite shows nontrivial value of polarization rotation angle caused by cosmic birefringence in the early universe. Moreover, the asymmetry of baryons versus anti-baryons still remains mysterious. Both of them indicates that there should be hidden new physics such as fundamental symmetry breaking. In this paper, we try to interpret these two events in framework of nonmetricity modified gravity. We introduce an interaction term between nonmetricity-based function and matter current, and calculate both the rotation angle and baryon-to-photon ratio. We also constrain the model parameters using the current observational data. With some specific examples, we demonstrate that in nonmetricity gravity theory, these two events can be interpreted in a unified way.  Nevertheless, the minimal coupling of nonmetricity scalar and the matter current might not be favored. 
\end{abstract}

\maketitle

\section{Introduction}
It is crucial to study the gravity theory that is governing the evolution our Universe. The Einstein Gravity, proposed by A. Einstein in 1915, is not only a simple and elegant theory, but also proved to be remarkably successful in explaining observations such as gravitational redshift, perihelion precession, bending of light and time delay, as well as predicting the gravitational waves \cite{Wald:1984rg}. Therefore, for long time it has been believed to be the right theory for describing gravity.  

However, the Einstein Gravity still has shortcomings. The most notorious one is its incompatibility with quantum theory, mainly due to its inability to be  renormalized \cite{tHooft:1974toh, Deser:1974cz, Deser:1974cy, Deser:1974zzd, Deser:1974nb, Deser:1974xq}. Moreover, phenomenologically, pure Einstein Gravity can neither drive cosmic acceleration (inflation) in the early universe, which is needed to solve the Big Bang problems \cite{Guth:1980zm, Linde:1981mu, Albrecht:1982wi, Starobinsky:1980te, Fang:1980wi, Sato:1981qmu}, nor can it act as dark sectors which plays critical roles in the late universe. These indicates that Einstein Gravity may be only effective rather than ultimate theory of gravity. Especially, at cosmological scales, the gravity may well go beyond the Einstein version.   

Among the modified gravity theories, one of special interest is the nonmetricity gravity \cite{Nester:1998mp, BeltranJimenez:2018vdo, BeltranJimenez:2019esp}. Belonging to the non-Riemann category of modified gravity theories, such a theory abandons the compatibility requirement of the metric, therefore the metric and the connection cannot be connected via Christoffel symbols, and become independent from each other. Although at linear level, this theory is equivalent to Einstein Gravity (a.k.a Symmetric Teleparallel Equivalence of General Relativity, STEGR \cite{Nester:1998mp, BeltranJimenez:2018vdo, BeltranJimenez:2019esp}), When going beyond linear level or coupled to other sectors, the two theories are quite different \cite{BeltranJimenez:2017tkd, Hu:2022anq, Hu:2023gui}. The differences allow us to apply such a theory into many aspects in cosmology that may or may not be done by Einstein Gravity itself, such as black hole physics \cite{Lin:2021uqa, DAmbrosio:2021zpm, Wang:2021zaz, Junior:2023qaq, Das:2024ytl}, inflation \cite{Capozziello:2022tvv, Capozziello:2024lsz}, bouncing cosmology \cite{Agrawal:2021rur, Mandal:2021wer, Hu:2023ndc, Sharif:2024yqj, Sharif:2024kby, Koussour:2024wtt, Shabani:2024ler}, and dark energy \cite{Koussour:2022ycn, Solanki:2022ccf, Lymperis:2022oyo, Yang:2024kdo, Solanki:2022rwu, Gadbail:2023suo, Bhardwaj:2023lph, Maurya:2024qtu, Basilakos:2025olm, Yang:2025mws}. See recent reviews \cite{Bahamonde:2021gfp, Heisenberg:2023lru}. Moreover, it is also interesting to place various observational constraints to test the signals induced by this theory \cite{Lazkoz:2019sjl, Barros:2020bgg, Anagnostopoulos:2021ydo, Atayde:2021pgb, Mandal:2021bpd, Maurya:2022vsn, DAgostino:2022tdk, Ferreira:2023awf, Mussatayeva:2023aoa, Shi:2023kvu, Bhagat:2023roh, Goswami:2023knh, Mandal:2023cag, Koussour:2023ulc, Shekh:2023rea, Oliveros:2023mwl, Javed:2024cqx, Yang:2024tkw, Gadbail:2024rpp, Koussour:2024sio, Wang:2024eai, Ge:2024tsx, Enkhili:2024dil, Su:2024avk, Myrzakulov:2024esv, Mhamdi:2024lxe, Wang:2024dkn, Maurya:2024chf, Shekh:2023zjc, Shukla:2024ocw, Paliathanasis:2025hjw, Boiza:2025xpn}.

The Planck data has released their new measurements of the cosmic birefringence angle $\Delta\chi$: $\Delta\chi=0.35\pm0.14$ deg ($68\%$ C.L.) (Data Release 3) \cite{Minami:2020odp} and $\Delta\chi=0.30\pm0.11$ deg ($68\%$ C.L.) (Data Release 4) \cite{Diego-Palazuelos:2022dsq}, excluding the trivial results of $\Delta\chi=0$ at more than $2\sigma$ \footnote{Recently, ACT has released their latest data on the cosmic birefringence angle $\Delta\chi=0.20\pm0.08$ deg(stat + optics)\cite{ACT:2025fju}, which is a $2.5\sigma$ departure from zero. Here we still consider data from Planck only.}. Since a non-zero birefringence angle will transfer parity-even CMB polarization patterns EE and BB to parity-odd polarization pattern EB, such results indicate that there should be some parity-violating process during the evolution of our Universe (See review \cite{Komatsu:2022nvu}). Moreover, data analysis from the BOSS galaxies survey also found evidence for parity violation at about $2.9\sigma$ \cite{Philcox:2022hkh}, consistent with the CMB data. Parity violation tends to imply new physics beyond Standard Model. In this work, we will first investigate how parity violation can be realized and constrained in nonmetricity gravity. As is well-known, a possible source of parity violation is the Chern-Simons term, where CMB photons, described by the electromagnetic field tensor $F_{\mu\nu}$ and its dual tensor $\tilde{F}_{\mu\nu}\equiv(1/2)\epsilon_{\mu\nu\rho\sigma}F^{\rho\sigma}$, couple to a dynamical scalar. Since $F_{\mu\nu}\tilde F^{\mu\nu}=\bf{E}\cdot\bf{B}$ violates the parity, such a term violates parity as well. The scalar can be quite arbitrary. In Einstein Gravity, it can be made of dark energy field \cite{Feng:2006dp, Li:2008tma}, axionlike field \cite{Capparelli:2019rtn, Murai:2022zur, Gonzalez:2022mcx, Eskilt:2023nxm}, or Ricci scalar $R$ \cite{Li:2004hh, Li:2006ss}, while in nonmetricity gravity, it can also be made of the nonmetricity scalar $Q$, which we will consider in the current context. 

Another interesting mystery is that why the amount of matter and anti-matter in our universe are not equivalent. It is an obvious fact since otherwise they will annihilate with each other and nothing could exist, but to be more precisely, the recent constraint on baryon asymmetry is given by $\eta\equiv n_B/n_\gamma \simeq (6.115\pm0.038)\times 10^{-10}$ (CMB+BBN+$Yp$+D, with $N_\nu=3$) \cite{ParticleDataGroup:2024cfk,Yeh:2024ors} (see detailed calculation of $\eta$ in \cite{Fields:2019pfx}). It is well-known that in standard case, the generation of baryon asymmetry should satisfy the three conditions proposed by Sakharov \cite{Sakharov:1967dj}: 1) the baryon number violation, 2) $C$ and $CP$ symmetry violation, and 3) out of thermal equilibrium. However, the third condition is actually established in $CPT$ symmetry conservation. In \cite{Cohen:1987vi} it was pointed out that when $CPT$ symmetry is violated, baryon asymmetry can be realized even in thermal equilibrium. By this means, people have proposed new types of baryogenesis mechanisms such as spontaneous baryogenesis \cite{Li:2001st, DeFelice:2002ir, Li:2002wd, Yamaguchi:2002vw, Chiba:2003vp, Alberghi:2003ws} and gravitational baryogenesis \cite{Davoudiasl:2004gf, Li:2004hh, Li:2006ss}. In the former, people use a scalar field (inflaton or dark energy field) derivatively coupled to a baryon current, and when the fields evolves into its vacuum expectation value which is non-zero, the $CPT$ violation is spontaneously broken. In the latter, the scalar field is replaced by some scalar function of gravity, e.g. $R$ or $f(R)$. Recently, people have extended the Riemannian gravity scalar to non-Riemannian ones, such as torsion \cite{Oikonomou:2016jjh}, nonmetricity \cite{LiMingZhe:2021vtu, Narawade:2024cgl, Samaddar:2024bnu, Alruwaili:2025gbw} and so on. 

These two events share some common properties, for example, both requires violation of fundamental discrete symmetries. This tempts people to think about their possible connections in physics. In this paper, we will take into account these two events together. Especially, with an interaction term of matter field and the nonmetricity scalar $Q$, we will see whether it is possible to interpret both of them in a unified way. The rest of the contents are arranged as follows: in Sec. \ref{sec2} we briefly introduce the nonmetricity gravity theory. In Sec. \ref{sec3} and \ref{sec4} we discuss about the cosmic birefringence and the baryogenesis in the nonmetricity-induced interaction term separately, and derive general formula of $\Delta\chi$ and $\eta$. In Sec. \ref{sec5} we provide some specific examples to see how constraints on $\Delta\chi$ and $\eta$ can be satisfied consistently. Sec. \ref{sec6} gives our conclusions and discussions.

\section{The nonmetricity gravity}
\label{sec2}
In nonmetricity gravity theory, the two basic elements-metric and connection, are viewed as independent variables. Therefore, rather than presented as Christoffel symbols only, the connection can be written as 
\begin{equation}
    \Gamma^{\alpha}_{\ \mu \nu}=\left\{ {}^{\, \alpha}_{\, \mu \nu} \right\}+L_{\  \mu \nu}^{\alpha}~,
    \label{connection}
\end{equation}
where $\left\{ {}^{\, \alpha}_{\,\mu \nu} \right\}$ is the usual Christoffel symbol: $\left\{ {}^{\, \alpha}_{\,\mu \nu} \right\}=g^{\alpha\lambda}(g_{\lambda\nu,\mu}+g_{\mu\lambda,\nu}-g_{\mu\nu,\lambda})/2$. Moreover, since this theory loses metric compatibility, the covariant derivative of metric $g_{\mu\nu}$ is no longer vanishing, namely
\begin{equation}
    \nabla_\alpha g_{\mu\nu}=\frac{\partial g_{\mu\nu}}{\partial x^\alpha}-g_{\nu\sigma}\Gamma^\sigma_{\ \mu\alpha}-g_{\sigma\mu}\Gamma^\sigma_{\ \nu\alpha}\equiv Q_{\alpha\mu\nu}~.
    \label{Qtensor}
\end{equation}
Thus $L_{\  \mu \nu}^{\alpha}$ can be presented as:
\begin{equation}
     L_{\  \mu \nu}^{\alpha}=- \frac{1}{2} g^{\alpha \lambda} \left( Q_{\mu \lambda \nu} + Q_{\nu \lambda \mu} - Q_{\lambda \mu \nu} \right)~.
\end{equation}

From the expression of connection \eqref{connection}, one can write down the Riemannian curvature tensor as 
\begin{eqnarray}
    R_{\ \beta \mu \nu}^{\alpha}(\Gamma)
&\equiv&\partial_{\mu} \hat\varGamma^{\alpha}_{\ \nu \beta}-\partial_{\nu} \hat\varGamma_{\ \mu \beta}^{\alpha}+\hat\varGamma_{\ \mu \lambda}^{\alpha} \hat\varGamma_{\ \nu \beta}^{\lambda}-\hat\varGamma_{\ \nu \lambda}^{\alpha} \hat\varGamma_{\ \mu \beta}^{\lambda}\nonumber\\
&=&\mathcal{R}_{\ \beta \mu \nu}^{\alpha}+ (\nabla_{\mu} L_{\ \nu \beta}^{\alpha}- \nabla_{\nu} L_{\ \mu \beta}^{\alpha}+L^{\alpha}_{\ \mu \lambda} L_{\ \nu \beta}^{\lambda}-L_{\ \nu \lambda}^{\alpha} L_{\ \mu \beta}^{\lambda})~,
\end{eqnarray}
while the Ricci Scalar turns out to be
\begin{equation}
    R=\mathcal{R}+(L^{\mu}_{\ \mu \lambda} L_{\ \nu}^{\lambda \ \nu}-L_{\ \nu\lambda }^{\mu} L_{\ \mu}^{\lambda\ \nu})+\nabla_{\mu} L^{\mu\  \nu}_{\ \nu}- \nabla_{\nu} L^{\mu\  \nu}_{\ \mu}~.
\end{equation}

Using the tensor $Q_{\alpha\mu\nu}$ defined in Eq. \eqref{Qtensor} and the general superpotential:
\begin{equation}
    P^{\alpha \mu \nu}
	=c_1Q^{\alpha \mu \nu}+c_2 Q^{\mu \alpha \nu}+c_3Q^{\alpha} g^{\mu \nu}+c_4 g^{\alpha \mu} \tilde{Q}^{\nu}+\frac{c_5}{2}(\tilde{Q}^{\alpha} g^{\mu \nu}+g^{\alpha \nu} Q^{\mu})~,
\end{equation}
it is also convenient to define a scalar form as: 
\begin{eqnarray}
    Q&=&Q_{\alpha\mu\nu}P^{\alpha \mu \nu} \nonumber\\
    &=&c_1Q_{\alpha\mu\nu}Q^{\alpha \mu \nu}+c_2 Q_{\alpha\mu\nu}Q^{\mu \alpha \nu}+c_3Q^{\alpha}Q_{\alpha}+c_4 \tilde{Q}^{\alpha}\tilde{Q}_{\alpha}+c_5Q^{\alpha}\tilde{Q}_{\alpha}~,
\end{eqnarray}
where $Q$ is called nonmetricity scalar, while
\begin{align}
    Q_{\alpha} \equiv Q_{\alpha}{ }_{\mu}{ }^{\mu}
    \, , \quad 
    \tilde{Q}_{\alpha} \equiv Q^{\mu}_{\ \mu \alpha}
    \, 
\end{align}
are two types of traces of the nonmetricity tensor $Q_{\alpha\mu\nu}$.

From \eqref{Qtensor} one can see that, the connection is still symmetric with respect to the subscript exchange: $\mu\leftrightarrow\nu$. This symmetry restrict connection from an arbitrary form. The simplest case is $\Gamma^{\alpha}_{\ \mu \nu}=0$ where the particle in this case moves as if it was in an inertia frame, so this is called coincidence gauge \cite{BeltranJimenez:2017tkd}. In this gauge, one has $\nabla\rightarrow\partial$, $L_{\  \mu \nu}^{\alpha}=-\left\{ {}^{\, \alpha}_{\,\mu \nu} \right\}$, and the Palatini formalism turns into metric formalism. Moreover, when one chooses $c_1=-1/4$, $c_2=1/2$, $c_3=1/4$, $c_4=0$, $c_5=-1/2$, then
\begin{eqnarray}
    Q&=& -\frac{1}{4}Q^{\alpha\mu\nu}Q_{\alpha\mu\nu}+\frac{1}{2}Q^{\nu\mu\alpha}Q_{\alpha\mu\nu}   
    +\frac{1}{4}Q^{\alpha}Q_{\alpha}  
    -\frac{1}{2}\tilde{Q}^{\alpha}Q_{\alpha} \nonumber\\
    &=& g^{\mu\nu}\left(L_{\ \sigma\mu}^{\alpha}L_{\ \nu \alpha}^{\sigma}-L_{\ \sigma\alpha}^{\alpha}L_{\ \mu\nu}^{\sigma}
    \right)    \nonumber\\
    &=& g^{\mu\nu}\left(
    \left\{ {}^{\, \alpha}_{\ \sigma \mu} \right\} 
    \left\{ {}^{\, \sigma}_{\ \nu \alpha}  \right\} 
    -  \left\{ {}^{\, \alpha}_{\ \sigma \alpha} \right\} 
    \left\{ {}^{\, \sigma}_{\ \mu \nu} \right\} 
    \right) \, .
\end{eqnarray}
This is analogy to the Einstein-Hilbert action in GR: 
\begin{equation}
\mathcal{R}=g^{\mu \nu}\left(\left\{ {}^{\,\alpha}_{\,\sigma \mu} \right\} \left\{ {}^{\,\sigma}_{\,\nu \alpha} \right\}-\left\{ {}^{\,\alpha}_{\,\sigma \alpha} \right\} \left\{ {}^{\,\sigma}_{\,\mu \nu} \right\}\right)+\nabla _{\alpha}\left[g^{\mu \nu}\left\{ {}^{\,\alpha}_{\,\mu\nu} \right\}-g^{\mu 
\alpha}\left\{ {}^{\,\nu}_{\,\mu\nu} \right\}\right]~,
\end{equation}
where the last term is total derivative. Therefore, if the action contains $Q$ only, it will be equivalent to the GR action moduli a total derivative, namely
\begin{equation}
    \mathcal{R}=Q-\nabla_{\alpha}\left(Q^{\alpha}-\tilde{Q}^{\alpha}\right)~.
\end{equation}
This is why such a theory is called STEGR. It is also obvious that when consider nonlinear to $Q$ such as $f(Q)$, it will be different from either $R$ or $f(R)$. 

However, one can see that the coincident gauge explicitly loses Lorentz symmetry. To restore symmetry, we consider more general connection: 
\begin{equation}
\Gamma^{\lambda}_{\,\mu \nu} = \frac{\partial x^{\lambda}}{\partial \xi^a}\frac{\partial^2 \xi^a}{\partial x^\mu\partial x^\nu}~,
\end{equation}
where the 4 St\"{u}eckelberg field $\xi^a$ transformed via metric $f_{ab}$, satisfying
\begin{equation}
g_{\mu\nu} = \frac{\partial\xi^{a}}{\partial x^{\nu}}\frac{\partial\xi^{b}}{\partial x^{\sigma}}f_{ab}~.
\end{equation}
When $\xi^a\rightarrow x^a$, this case reduce to that of coincidence gauge.

\section{parity violation and cosmic birefringence}
\label{sec3}
\subsection{General Formalism}
Let's consider the nonmetricity gravity theory which contains a Chern-Simons-like nonminimal coupling between gravity and the electromagnetic field. The action is
\begin{align}
\mathcal{S} =-\frac{1}{2\kappa}\int_{}{\mathrm{d}^4x}\sqrt{-g}f(Q)+\frac{\alpha}{M^2} \int{d^4}x\sqrt{-g}l(Q)\tilde{F}^{\mu \nu}F_{\mu \nu}+\frac{1}{4}\int{d^4}x\sqrt{-g}F^{\mu \nu}F_{\mu \nu}~,
\end{align}
where the gravity part is a general function of $Q$. The second term is expected to violate the parity symmetry, in which $l(Q)$ is the coupling function. The magnitude of this term is described by the parameter $\alpha$. The last term is the kinetic term of the electromagnetic field standing for the CMB photons. 

It is straightforward to obtain equations of motion that govern the evolution of this system. Since the nonmetricity gravity theory has two independent variables, namely the metric and the connection, therefore, varying the action with the metric, one gets the metric field equation:
\begin{eqnarray}
&&-2\nabla_{\alpha}
\left[ \left( f_Q{ -2\kappa \frac{\alpha}{M^2} l_Q\tilde{F}^{ab}F_{ab}} \right) {P^{\alpha}}_{\mu \nu} \right] +\left( f_Q{ -2\kappa \frac{\alpha}{M^2} l_Q\tilde{F}^{ab}F_{ab}} \right) \left( {L^{\alpha}}_{\beta \nu}{L^{\beta}}_{\mu \alpha}-{L^{\beta}}_{\alpha \beta}{L^{\alpha}}_{\mu \nu} \right)  \nonumber\\
&&-\frac{1}{2}g_{\mu \nu}f-8{\kappa \frac{\alpha}{M^2} \epsilon{_{\mu}}^{\lambda \alpha \beta}F_{\nu \lambda}F_{\alpha \beta}l(Q)}=\kappa \left( F_{\mu \alpha}{F_{\nu}}^{\alpha}-\frac{1}{4}g_{\mu \nu}F^{\alpha \beta}F_{\alpha \beta} \right)~, 
\end{eqnarray}
where $l_Q\equiv dl(Q)/dQ$. Varying the action with the connection, one gets the connection field equation:
\begin{equation}
\hat{\nabla}_{\nu}\hat{\nabla}_{\mu}\left[ \sqrt{-g}\left( f_Q{-2\kappa \frac{\alpha}{M^2}l_Q \tilde{F}^{ab}F_{ab}} \right) P_{\,\,    \alpha}^{\,\,\nu \mu} \right] =0~.
\end{equation}


Moreover, the coupling term can also be rewritten into the form:
\begin{align}
 \int{d^4}x\sqrt{g}l(Q)\tilde{F}^{\mu \nu}F_{\mu \nu}=2 \int{d^4}x\sqrt{g}l_Q\partial _{\mu}Q\tilde{F}^{\mu \nu}A_{\nu}~,
\label{csterm}
\end{align}
and varying the action with respect to the electromagnetic field $A_\nu$, one gets:
\begin{eqnarray}
\delta S&=&\frac{2\alpha}{M^2} \int{d^4}x\sqrt{g}l_Q\left( \partial _{\mu}Q \right) \delta A_{\nu}\widetilde{F}^{\mu \nu}+\frac{1}{2}\int{d^4}x\sqrt{g}F^{\mu \nu}\left( \partial _{\mu}\delta A_{\nu}-\partial _{\nu}\delta A_{\mu} \right) \nonumber\\
&&+\frac{4\alpha}{M^2} \int{d^4}x\sqrt{g}\left( l_Q\partial _{\alpha}Q \right) A_{\beta}\epsilon ^{\mu \nu \alpha \beta}\left( \partial _{\mu}\delta A_{\nu} \right) 
\\
&=&\int{d^4}x\left( l_Q\frac{4\sqrt{g}\alpha}{M^2} \left( \partial _{\mu}Q \right) \widetilde{F}^{\mu \nu}-\sqrt{g}\left( \nabla _{\mu}F^{\mu \nu} \right) \right) \delta A_{\nu}
= 0~,
\end{eqnarray}
from which we get the modified electromagnetic field equation: 
\begin{align}
\nabla _{\mu}F^{\mu \nu}=&\frac{4\alpha}{M^2} l_Q\left( \partial _{\mu}Q \right) \widetilde{F}^{\mu \nu}~\label{equation A 1}~,
\\
\nabla _{\nu}\widetilde{F}^{\mu \nu}=&0~.\label{equation A 2}
\end{align}

It is also easy to check the properties of each element in Eq. \eqref{csterm} under the transformation of charge conjugation ($C$), parity ($P$) and time reversal ($T$), which are listed in Table \ref{CPT} below. From this table one can see that, this term violates parity symmetry explicitly, which may cause the cosmic birefringence in CMB. 
\begin{table}[htbp]
\centering
\begin{tabular}{|c|c|c|c|c|}
\hline
	&$A_{\nu}$&$\partial _{\mu}l(Q)$&${F}^{\mu \nu}$&$\epsilon ^{\mu \nu \alpha \beta}$	\\
    \hline
	\makecell{Charge\\ Conjugation}&$A_{\nu}\xrightarrow{C}-A_{\nu}$&$\partial _{\mu}l(Q)\xrightarrow{C}\partial _{\mu}l(Q)$&$F_{\alpha \beta}\xrightarrow{C}-F_{\alpha \beta}$&$\epsilon ^{\mu \nu \alpha \beta}\xrightarrow{C}\epsilon ^{\mu \nu \alpha \beta}$	\\
    \hline
	Parity&\makecell{$A_0\left( x \right) \xrightarrow{P}A_0\left( -x \right)$\\$A_i\left( x \right) \xrightarrow{P}-A_i\left( -x \right)$}&\makecell{$\partial _0l(Q\left( x \right)) \xrightarrow{P}\partial _0l(Q\left( -x \right)) $\\$\partial _il(Q\left( x \right)) \xrightarrow{P}-\partial _il(Q\left(- x \right))$} 
&\makecell{$F_{0j}\left( x \right)\xrightarrow{P}-F_{0j}\left( -x \right)$\\$ F_{ij}\left( x \right)\xrightarrow{P}F_{ij}\left( -x \right)$} &$\epsilon^{\mu \nu \alpha \beta}\xrightarrow{P}-\epsilon ^{\mu \nu \alpha \beta}$	\\
    \hline
	\makecell{Time\\ Reversal}&\makecell{$A_0\left( t \right) \xrightarrow{T}A_0\left( -t \right) $\\$A_i\left( t \right) \xrightarrow{T}-A_i\left( -t \right)$}&\makecell{$\partial _0l(Q\left( t \right)) \xrightarrow{T}-\partial _0l(Q\left( -t \right))$\\$\partial _il(Q\left( t \right)) \xrightarrow{T}\partial _il(Q\left( -t \right))$ 
}	&\makecell{$F_{0j}\left( t \right)\xrightarrow{T}F_{0j}\left( -t \right)$\\ $F_{ij}\left( t \right)\xrightarrow{T}-F_{ij}\left( -t \right)$
} &$\epsilon ^{\mu \nu \alpha \beta}\xrightarrow{T}-\epsilon ^{\mu \nu \alpha \beta}$	\\
    \hline
\end{tabular}
\caption{The properties of each component in Chern-Simons-like term \eqref{csterm} under $C$, $P$ and $T$ transformations.}
\label{CPT}
\end{table}

\subsection{The Geometric Optics Approximation}
In the case where the typical wavelength of the CMB photons $\lambda$ is much smaller than the radius of curvature $r_c$ of spacetime through which the photons are propagating, it is suitable to make use of the ``geometric optics approximation" \cite{Misner:1973prb} to solve Eqs. \eqref{equation A 1} and \eqref{equation A 2} \cite{Li:2006ss, Li:2008tma}. Under this approach, the electromagnetic field vector can be formulated as a perturbative expansion: 
\begin{equation}
    A_\mu=\text{Re}[(a_\mu+\epsilon b_\mu+\epsilon^2 c_\mu+\cdots)e^{iS/\epsilon}]~,
\end{equation}
where the expansion parameter $\epsilon\equiv\lambda/(2\pi r_c)$. Moreover, one can define the wave vector as the gradient of the phase function:  
\begin{align}
k_\mu \equiv \nabla_\mu S~,
\end{align}
which characterizes the photon’s propagation direction. Therefore, in the case where the phase varies much faster than the amplitude, one has:
\begin{equation}
    \nabla_\mu A_\nu=\nabla_\mu\text{Re}[(a_\nu+\epsilon b_\nu+\epsilon^2 c_\nu+\cdots)e^{iS/\epsilon}]\sim\text{Re}\left[\frac{ik_\mu}{\epsilon}(a_\nu+\epsilon b_\nu+\epsilon^2 c_\nu+\cdots)e^{iS/\epsilon}\right]~.
\end{equation}
Furthermore, the electromagnetic field tensor, $F_{\mu\nu}\equiv\nabla_\mu A_\nu-\nabla_\nu A_\mu$, can be formulated as:
\begin{equation}
F_{\mu\nu} = \text{Re}\left[\frac{i}{\epsilon}(a_{\mu\nu} + \epsilon b_{\mu\nu} + \epsilon^2 c_{\mu\nu} + \dots) e^{iS/\epsilon}\right]\label{equation expansion F}~,
\end{equation} 
the amplitude of which satisfies the relation:
\begin{equation}
    a_{\mu\nu}=k_\mu a_\nu-k_\nu a_\mu~,~b_{\mu\nu}=k_\mu b_\nu-k_\nu b_\mu~,~c_{\mu\nu}=k_\mu c_\nu-k_\nu c_\mu~,\cdots
\end{equation}

Combining Eqs. \eqref{equation A 1} and \eqref{equation A 2} one gets the following equation:
\begin{align}
\Box F_{\rho \sigma}-\frac{4\alpha}{M^2}\nabla_{\rho}[ l_Q\left( \partial _{\mu}Q \right) \widetilde{F}_{\,\, \sigma}^{\mu}]+\frac{4\alpha}{M^2}\nabla_{\sigma}[ l_Q \left( \partial _{\mu}Q \right) \widetilde{F}_{\,\, \rho}^{\mu}]-[F_{\,\,\rho}^{\alpha}R_{\alpha \sigma}-F_{\,\,\sigma}^{\alpha}R_{\alpha \rho}-F^{\mu \alpha}R_{\alpha \mu \rho \sigma}]=0~.
\label{equation A 3} 
\end{align}

Substituting ~\eqref{equation expansion F} into \eqref{equation A 3} one can get the equations in each order of $\epsilon$, which has already been discussed in \cite{Li:2006ss, Li:2008tma}. Note that each derivative operator will generate a factor of $i/\epsilon$, therefore it is easy to observe that the first term in \eqref{equation A 3}, $\Box F_{\rho \sigma}\sim1/\epsilon^2$, is the leading order in $\epsilon$, while the second two terms, $(4\alpha/M^2)\{-\nabla _{\rho}[ l_Q\left( \partial _{\mu}Q \right) \widetilde{F}_{\,\, \sigma}^{\mu}]+\nabla _{\sigma} [ l_Q\left( \partial _{\mu}Q \right) \widetilde{F}_{\,\, \rho}^{\mu}]\}\sim 1/\epsilon$, are the next leading order. The terms containing Ricci and Riemann tensors are last leading order terms, and thus can be neglected. 
%
We only consider the $1/\epsilon^2$ and $1/\epsilon$ orders. At the order of $1/\epsilon^2$, Eqs.~\eqref{equation A 3} becomes:
\begin{equation}
k_{\mu}k^{\mu}=0~,
\label{k=0}
\end{equation}
which means the CMB photons keep zero rest mass,
while at the order of $1/\epsilon$, Eq.~\eqref{equation A 3} turns out to be:
\begin{equation}
\mathcal{D} a^{\nu}+\frac{\theta}{2}a^{\nu}=\frac{2\alpha}{M^2} l_Q\left( \partial _{\mu}Q \right) \epsilon ^{\mu \nu \rho \sigma}k_{\rho}a_{\sigma}~,
\end{equation}
where ${\cal D}\equiv k^\mu\nabla_\mu$, $\theta\equiv\nabla_\mu k^\mu$. One can also express $a^{\nu}$ as the combination of a scalar magnitude $A$ and a unit-normalized polarization vector $\zeta^{\nu}$:
\begin{equation}
a^{\nu}=A\zeta^{\nu}~,
\end{equation}
then the equation can be expressed by unit-normalized polarization vector $\zeta^{\nu}$:
\begin{equation}
k^{\mu}\nabla _{\mu}\zeta ^{\nu}=\frac{2\alpha}{M^2} l_Q\left( \partial _{\mu}Q \right) \epsilon ^{\mu \nu \rho \sigma}k_{\rho}\zeta_{\sigma}~.
\label{zetaeq}
\end{equation}
Note that the covariant derivative appearing here corresponds to Christoffel symbols.

\subsection{Cosmic Birefringence and polarization angle}
The Levi-Civita symbol $\epsilon ^{\mu \nu \rho \sigma}$ in Eq. \eqref{zetaeq} will cause component mixing in the polarization vector which leads to polarization angle. To see this, let's work in a spatially flat, homogeneous, and isotropic universe which is described by the Friedmann–Lemaître–Robertson–Walker metric:
\begin{equation}
    ds^2=-dt^2+a^2(t)\delta _{ij}dx^idx^j~,
\end{equation}
where $a(t)$ is the scale factor. From this metric we gets the non-zero components of Christoffel symbols as: 
\begin{equation}
    \Gamma^i_{0j}=H\delta^i_j~,~\Gamma^0_{ij}=a^2H\delta_{ij}~.
\end{equation}



Assuming the electromagnetic waves are propagating along the $k^3$-axis direction, the wave vector $k^{\mu}=(k^0,0, 0, k^3)$, and Eq. \eqref{k=0} gives rise to $(k^0)^2=a^2(k^3)^2$. Then the two independent polarization vector components are $\zeta ^1$,$\zeta^2$, respectively. From Eq. \eqref{zetaeq} we have:
\begin{align}
\frac{d\zeta^1}{d\lambda}+Hk^0\zeta^1=& \frac{2\alpha}{M^2}l_Q\dot{Q}ak^3\zeta^2~,
\\
\frac{d\zeta^2}{d\lambda}+Hk^0\zeta^2=&-\frac{2\alpha}{M^2}l_Q\dot{Q}ak^3\zeta^1~.
\end{align}

The polarization angle is defined as $\chi \equiv \mathrm{arc}\tan \left( \zeta^2/\zeta^1 \right)$. When the polarization vector propagates from redshift $z_{rec}$ of recombination time to now ($z=0$), the polarization angle is rotated by
\begin{equation}
\label{deltachi}
    \Delta\chi=\int^{z_{rec}}_{0} d\chi =\frac{2\alpha}{M^2}\int^{z_{rec}}_{0}l_QdQ=\frac{2\alpha}{M^2}[l(Q(z_{rec}))-l(Q(0))]~.
\end{equation}



Such a polarization angle will in turn cause the mixing of correlations in CMB power spectra, and produce parity-odd correlations such as $TB$ and $EB$ via the following equations \cite{Lue:1998mq, Feng:2006dp, Li:2006ss, Li:2008tma}: 
\begin{equation}
    {C'}_l^{TE}=C_l^{TE}\cos(2\Delta\chi)~,~{C'}_l^{TB}=C_l^{TE}\sin(2\Delta\chi)~,
\end{equation}
as well as other modifications of correlations:
\begin{eqnarray}
    {C'}_l^{EE}&=&C_l^{EE}\cos^2(2\Delta\chi)+C_l^{BB}\sin^2(2\Delta\chi)~,\\
    {C'}_l^{BB}&=&C_l^{EE}\sin^2(2\Delta\chi)+C_l^{BB}\cos^2(2\Delta\chi)~,\\
    {C'}_l^{EB}&=&\frac{1}{2}(C_l^{EE}-C_l^{BB})\sin(4\Delta\chi)~.
\end{eqnarray}
Using the polarization data from Planck 2018, nowadays putting limits on the rotation of the polarization plane $\Delta \chi =0.30\pm 0.11 \deg$ ($68\%$ C.L.) \cite{Diego-Palazuelos:2022dsq}.

\section{CPT violation and baryogenesis}
\label{sec4}

In this section, we consider the baryogenesis mechanism in our model and its constraints by observations. We consider the coupling of the nonmetricity term to baryon current in the form of:
\begin{align}
\label{baryointeract}
L_{eff}=-\frac{\alpha}{M_{}^{2}}\partial _{\mu}l(Q)J^{\mu}~,
\end{align}
where $\alpha$ is dimensionless coupling constant, while $M$ is the cutoff mass scale. As has been mentioned in the introduction, usually baryogenesis requires the three Sakharov conditions, however, when the interaction term violates $CPT$ symmetry, baryogenesis can also take place. In the case of Eq. \eqref{baryointeract}, when $\partial _{\mu}l(Q)$ gets a non-zero vacuum expectation value in a homogeneous cosmic background, namely $\langle\partial _{\mu}l(Q)\rangle=\langle l_Q\dot{Q}\rangle\neq0$, this term will violate the $CPT$ symmetry. Note that this is somehow difficult in the Ricci-scalar-based interaction term, because $\dot R=0$ in the radiation-dominant age when baryogenesis is expected to happen, unless one modifies $R$ into an function $l(R)$ to ensure $l_R\dot{R}\neq 0$ \cite{Li:2004hh, Li:2006ss} (see also \cite{Davoudiasl:2004gf} for discussion on other possibilities). However, for the nonmetricity-based term, we don't need to worry about this since $\dot Q\neq 0$, and this can be seen as an advantage of $Q$ over $R$ in this issue. 

A non-vanishing $l_Q\dot{Q}$ provides baryons/antibaryons with an effective chemical potential as:
\begin{align}
\mu _b=-\mu _{\bar{b}}\sim -\frac{\alpha l_Q\dot{Q}}{M^2}~,
\end{align}
which leads to a violation of the baryon-antibaryon number balance under thermal equilibrium, giving rise to B-violation. In the high-temperature limit, the expression takes the form:
\begin{align}
n_b-n_{\bar{b}}\simeq -\frac{g_bT^3}{6\pi ^2}\left( \frac{\pi ^2\mu _B}{T}+\left( \frac{\mu _B}{T} \right) ^3 \right) \simeq -\left. \frac{g_b\alpha l_Q\dot{Q}T^2}{3M_{}^{2}} \right|_{T_D}~,
\end{align}
where $g_b=2$ is the total number of intrinsic degrees of freedom of baryons and $T_D$ is the decoupling temperature. Below this temperature, as the universe expands and gets cool, the asymmetry will remain. On the other hand,
during the radiation-dominated era of the universe, the expression for photon number density is given by:
\begin{align}
    n_\gamma=\frac{\zeta(3)}{\pi^2}g_bT^3~,
\end{align}
here $\zeta(3)\simeq 1.20206\cdots$ is the Riemann zeta function of $3$. 
Therefore, the baryon-to-photon ratio is:
\begin{align}
\eta=\frac{n_b-n_{\bar{b}}}{n_\gamma}=-\frac{\pi^2}{3\zeta(3)}\frac{\alpha l_Q\dot{Q}}{M^{2}T_D}~.
\end{align}

Neglecting the influence of coupling terms on the Hubble parameter within the framework of $f(Q)$ gravity theory, the expression for the Hubble parameter is formulated as follows\cite{Beh_2022}:
\begin{align}
\label{hubble}
H^2=\frac{H_{0}^{2}}{f_Q}[\Omega_{m0}(1+z)^3+\Omega_{r0}(1+z)^4+\Omega _{DE}]~,
\end{align}
where
\begin{align}
\Omega _{DE}:=\frac{1}{H_{0}^{2}}\left( \frac{f_QQ}{6}-\frac{f}{6} \right)~. 
\end{align}
During the radiation-dominated era, one has $H\simeq 5f^{-1/2}_Q g_{*}^{1/2}T^2/(3M_p)$ as well as $Q=6H^2$. $g_\ast$ is the total number of effectively massless degrees of freedom, with $g_\ast\simeq106.75$ for the temperature of the universe higher than $O(100)\rm{GeV}$. Then the baryon number asymmetry becomes:
\begin{align}
    \eta\simeq 3.398\times10^4\pi^2\alpha f^{-3/2}_Q l_Q\left(\frac{g_\ast|_{T_D}}{106.75}\right)^\frac{3}{2}\left(\frac{T_D}{M_p}\right)^3\left(\frac{T_D}{M}\right)^2~.
\end{align}
As mentioned in the introduction, the current constraint on $\eta$ by CMB+BBN+$Yp$+D (with $N_\nu=3$) turns out to be $\eta \simeq (6.115\pm0.038)\times 10^{-10}$ \cite{ParticleDataGroup:2024cfk,Yeh:2024ors}. 

\section{Specific examples}
\label{sec5}
In this section, we consider some specific examples that can account for both processes discussed above. We assume that at least in the early universe when these processes took place, the $f(Q)$ gravity still behaves like the STEGR, namely $f_Q\simeq 1$. Therefore, we mainly consider the impact of the nontrivial coupling function $l(Q)$.

\subsection{Case I: $l(Q)=Q$}
This is the simplest case, which we may call as the ``minimal coupling". In this case, the birefringence rotation angle $\Delta\chi$ in Eq. \eqref{deltachi} becomes:
\begin{equation}
    \Delta\chi=\frac{2\alpha}{M^2}[Q(z_{rec})-Q(0)]~.
\end{equation}

Making use of Eq. \eqref{hubble}, with values $c=3\times 10^8\rm{m/s}$, $\Omega_{m0}\simeq 0.315$, $\Omega_{r0}\simeq 9.2\times 10^{-5}$, $z_{rec}\simeq1100$, $H_0\simeq70\rm{km/s/Mpc}$ \footnote{There are uncertainties on value of $H_0$ which is known as the notorious ``Hubble tension". However, in the current theme we don't bother about the exact number, and just take the mean value of $H_0$ for discussion.}, $M\simeq1.2\times10^{19}\rm{GeV}$, one gets:
\begin{align}
\Delta \chi=&\frac{12\alpha}{M^2}[H^2(1100)-H^2(0)]=1.03\times10^{-112}\alpha~,
\end{align}
while the constraint on $\Delta\chi$ gives:
\begin{align}
\label{conbire1}
3.21\times 10^{109}<&\alpha <6.93\times 10^{109}~.
\end{align}
On the other hand, the constraints on baryon asymmetry gives
\begin{align}
    \eta \simeq 3.398\times10^4\pi^2\alpha \left(\frac{g_\ast|_{T_D}}{106.75}\right)^\frac{3}{2}\left(\frac{T_D}{M_p}\right)^3\left(\frac{T_D}{M}\right)^2~.
\end{align}
Considering the common case that the freeze-out temperature should be above the nucleon freeze-out scale yet remain well below the Grand Unification Theory (GUT) scale, namely $10\text{MeV}<T_D<10^{12}\text{GeV}$, the constraint on $\eta$ requires the allowable range for parameter to be within $4.509\times 10^{20}<\alpha <4.566\times10^{90}$. This clearly contradicts with the requirement from the birefringence \eqref{conbire1}. Therefore, to reconcile between the two constraints involves a severe fine-tuning problem.

\subsection{Case II: $l(Q)=M^2\ln (Q/M_p^2)$}
In order to reconcile the constraints from the two processes, now we consider a nontrivial function of $l(Q)$. The first example is $l(Q)=M^2\ln(Q/M_p^2)$, while the same function form of Ricci scalar $R$ has actually been considered in gravitational baryogenesis in \cite{Li:2004hh,Li:2006ss}. Thanks to the logarithm function, which can alleviate the huge hierarchy between $Q$ and $M_p^2$ (or the cutoff scale $M^2$) to a milder level, the constraint on the coefficient $\alpha$ could be made natural. In this case, the birefringence rotation angle then becomes:
\begin{align}
\Delta \chi=2\alpha [\ln(Q(z_{rec})/M_p^2)-\ln(Q(0)/M_p^2)]\simeq 40.21\alpha~,
\end{align}
according to which we have 
\begin{align}
\label{conbire2}
8.23\times10^{-5}<\alpha<1.78\times10^{-4}~.
\end{align}
Moreover, the baryon number asymmetry is
\begin{align}
\eta \simeq 19.10\pi^2\alpha\left(\frac{g_\ast}{106.75}\right)^\frac{1}{2}\frac{T_D}{M_p}~.
\end{align}
For $10\text{MeV}<T_D<10^{12}\text{GeV}$, We find that the constraint on $\eta$ allows $\alpha$ to vary within $3.868\times10^{-5}<\alpha<3.917\times10^{11}$. This is obviously consistent with that from constraint \eqref{conbire2}. Especially, if we choose $\alpha\sim 10^{-4}$ according to \eqref{conbire2}, we find the freeze-out temperature to be $T_D\simeq10^{12}\text{GeV} $, which is below the GUT scale.

\subsection{Case III: $l(Q)=-M^2\sinh^{-1}\left(\sqrt{1+M_p^2/Q}\right)$}
Another example is that $l(Q)$ could be made of a more complicated function, such as $l(Q)=-M^2\sinh^{-1}\left(\sqrt{1+M_p^2/Q}\right)$. Being a function which approaches to a flatter value when $Q$ goes smaller ($M_p^2/Q\rightarrow\infty$), it could also alleviate the hierarchy between $Q$ and $M_p^2$. In this case, the birefringence rotation angle turns out to be:
\begin{align}
\Delta \chi=-2\alpha \left[\sinh^{-1}\left(\sqrt{1+M_p^2/Q(z_{rec})}\right)-\sinh^{-1}\left(\sqrt{1+M_p^2/Q(0)}\right)\right]\simeq 20.11\alpha~,
\end{align}
which give rise to
\begin{align}
\label{conbire3}
1.65\times10^{-4}<\alpha<3.56\times10^{-4}~.
\end{align}
On the other hand, The baryon number asymmetry is:
\begin{align}
\eta\simeq 9.56\pi^2\alpha\left(\frac{g_\ast}{106.75}\right)^\frac{1}{2}\frac{T_D}{M_p}~.
\end{align}
For $10\text{MeV}<T_D<10^{12}\text{GeV}$, the constraint on $\eta$ leads to $7.737\times10^{-5}<\alpha<7.834\times10^{11}$. This is also consistent with that from constraint \eqref{conbire3}. Similar to the previous case, for $\alpha\sim 10^{-4}$ given by \eqref{conbire3}, the freeze-out temperature is $T_D\simeq10^{12}\text{GeV}$ as well.
\begin{figure}[H]
    \centering \includegraphics[width=0.8\linewidth]{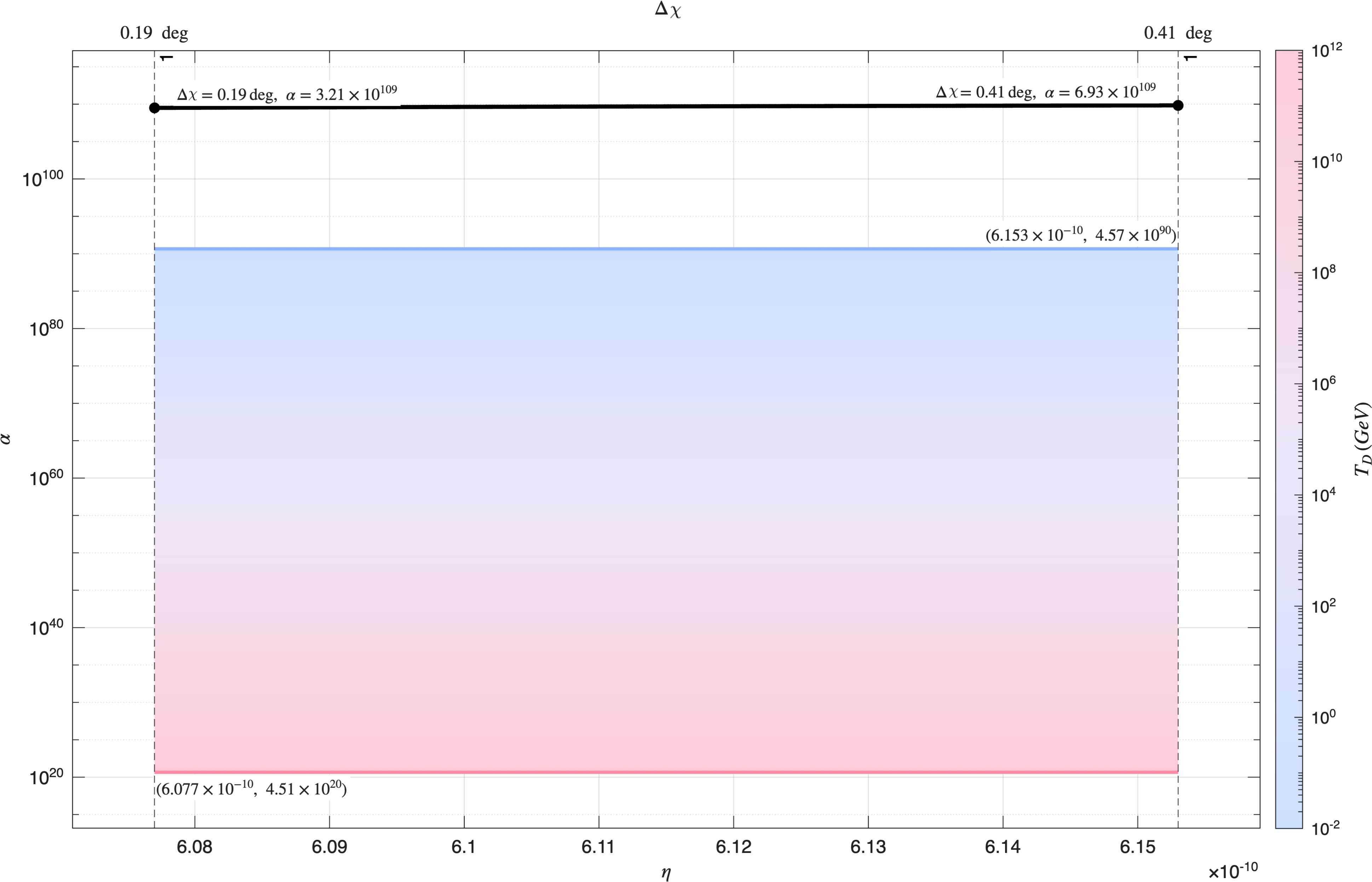}
    \caption{The parameter $\alpha$ versus the rotation angle $\Delta\chi$ and baryon asymmetry $\eta$ in case I. The region between dashed lines are the region allowed by the observations at $1\sigma$ level. The black line and the color band are for the constraints on $\alpha$ by $\Delta\chi$ and by $\eta$, respectively.}
    \label{fig:caseI}
\end{figure}

\begin{figure}[H]
    \centering \includegraphics[width=0.8\linewidth]{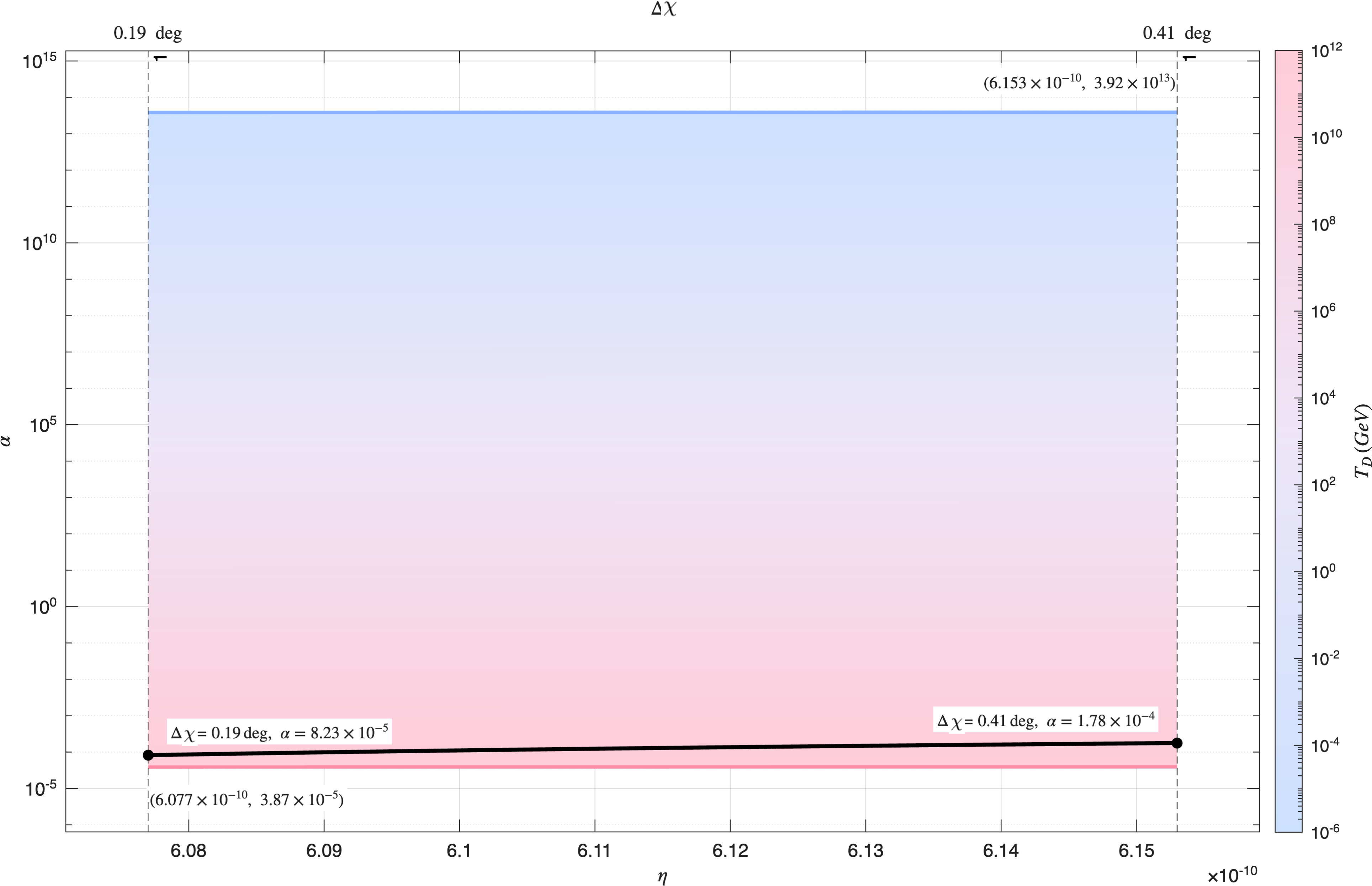}
    \caption{The parameter $\alpha$ versus the rotation angle $\Delta\chi$ and baryon asymmetry $\eta$ in case II. The dashed lines, The black line and the color band denotes the same things as in Fig. \ref{fig:caseI}.}
    \label{fig:caseII}
\end{figure}

\begin{figure}[H]
    \centering 
    \includegraphics[width=0.8\linewidth]{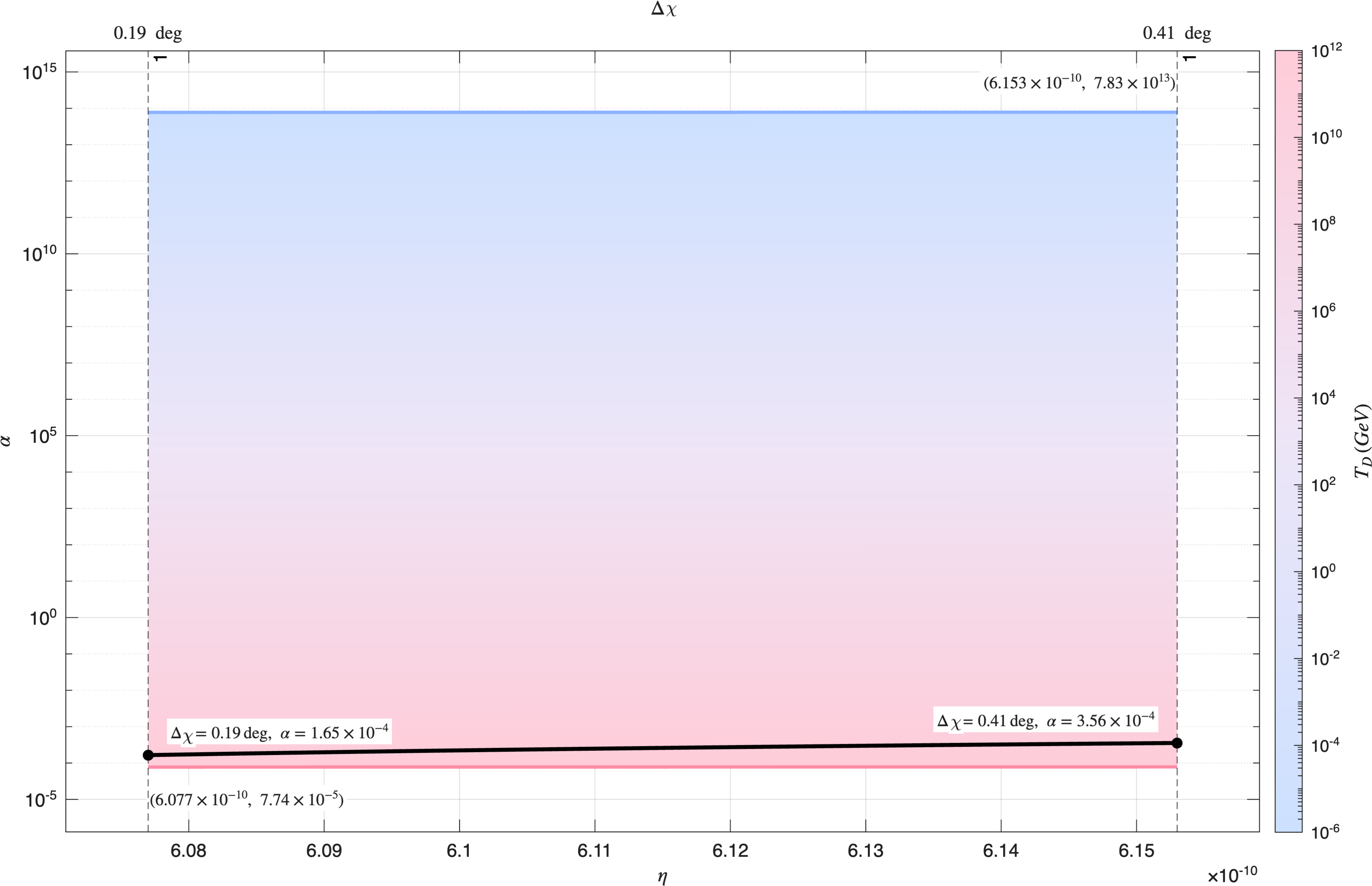}
    \caption{The parameter $\alpha$ versus the rotation angle $\Delta\chi$ and baryon asymmetry $\eta$ in case III. The dashed lines, The black line and the color band denotes the same things as in Fig. \ref{fig:caseI}.}
    \label{fig:caseIII}
\end{figure}

{In Figs. \ref{fig:caseI}, \ref{fig:caseII} and \ref{fig:caseIII}, we plot the constraints on $\alpha$ from the allowed parameter space of both $\Delta\chi$ and $\eta$ in all the three examples, respectively. In the figures, we see that in case I, the constraint on $\alpha$ by $\Delta\chi$ and $\eta$ cannot be overlapped, meaning that we cannot satisfy the two constraints simultaneously, while in case II and case III we can. This indicates that while one can interpret both the two events in a unified way in framework of nonmetricity gravity, we may need a more complicated coupling form than the minimal one. }
\section{summary and discussion}
\label{sec6}
In this paper, we consider the application of nonmetricity gravity to two important cosmological events, namely cosmic birefringence and baryon asymmetry, in the early universe. Both of them implies fundamental symmetry breaking, such as parity or $CPT$ violation. Via an interaction term with a general function of the nonmetricity scalar $Q$ coupled to a matter current, both events can be naturally interpreted. General expressions for polarization angle $\Delta\chi$ (for cosmic birefringence) as well as baryon-to-photon ratio $\eta$ (for baryon asymmetry) are also obtained. 

With three explicit examples, we try to see whether our model could also fit the observational constraints, which have been updated in recent years. For the simplest function of $l(Q)=Q$, we find that it is difficult to satisfy the two constraints simultaneously, unless the model parameters are unnaturally tuned. This is because the hierarchy between $Q$ and the cutoff scale makes the parameter $\alpha$ extremely large. However, for other choices such as $l(Q)=M^2\ln (Q/M_p^2)$ and $l(Q)=-M^2\sinh^{-1}\left(\sqrt{1+M_p^2/Q}\right)$ where the functions can alleviate the hierarchy to an acceptable level, our model can fit the observational data very well. {This means that in order to interpret both the two events in the framework of nonmetricity gravity theory, a more complicated coupling form to the CMB photons/baryon current will be favored than the minimal coupling form. }

Nonmetricity modified gravity theory can be applied to many aspects in cosmology. The most popular applications may be in inflation or dark energy era, with many theoretical studies and observational constraints done. Nonetheless, our study focuses on another less studied but important era, namely the early (but not that early) universe from $0.1\text{eV}<T<10^{12}\text{GeV}$. We hope our work can contribute to the construction and global tests of the  nonmetricity gravity cosmology. Besides playing role in these events, nonmetricity may also have impacts on following physical processes, such as formation/evolution of cosmic foreground, galaxies and dark matter.

In nonmetricity modified gravity theories, there could also be other kind of parity-violating terms. For example, the nonmetricity tensor coupled with scalar field can give rise to parity violation in gravity sector \cite{Conroy:2019ibo, Li:2022vtn, Chen:2022wtz, Zhang:2023scq, Califano:2023aji}, just like the Chern-Simons term, which can be constrained by the measurements of gravitational waves propagation \cite{LIGOScientific:2017vwq, LIGOScientific:2017zic} (see also other forms in e.g. \cite{Iosifidis:2020dck, Nojiri:2025jhh}). These works indicate that there is plentiful information about new physics hidden in this kind of gravity theory, which is to be explored in future works.

\begin{acknowledgments}
We thank Kun Hu, Mingzhe Li, Wei Chao, Yu-Cheng Qiu, Tao Liu for useful discussions. This work is supported by the National Key Research and Development Program of China (Grant No. 2021YFC2203100). 
\end{acknowledgments}

\bibliographystyle{apsrev4-1}
\bibliography{References}
\end{document}